\documentclass[reprint, amsmath,amssymb,aps,]{revtex4-2}
 
\usepackage{mathbbol}
\usepackage{graphicx,color}
\usepackage{amsmath,amssymb,bm,amsthm,amstext}
\usepackage{mathtools}
\usepackage{dsfont}
\usepackage{simplewick}
\usepackage{braket}
\usepackage{qcircuit}
\usepackage{array}
\usepackage{soul}
\usepackage{dcolumn}
\usepackage[scr]{rsfso}

\begin{document}

\pagecolor{white}

\preprint{APS/123-QED}

\title{Nuclear Induction Lineshape: Non-Markovian Diffusion with Boundaries}

\thanks{Correspondence: lsbouchard@ucla.edu (L.-S. B.)}%

\author{Mohamad Niknam}
\affiliation{Department of Chemistry and Biochemistry, University of California Los Angeles, 607 Charles E. Young Drive East, Los Angeles, CA 90095-1059, USA}
\affiliation{Center for Quantum Science and Engineering, UCLA}
\author{Louis-S. Bouchard*}
\affiliation{Department of Chemistry and Biochemistry, University of California Los Angeles, 607 Charles E. Young Drive East, Los Angeles, CA 90095-1059, USA}
\affiliation{Center for Quantum Science and Engineering, UCLA}
\email{lsbouchard@ucla.edu}%

\date{\today}

\begin{abstract}
The dynamics of viscoelastic fluids are governed by a memory function, essential yet challenging to compute, especially when diffusion faces boundary restrictions. We propose a computational method that captures memory effects by analyzing the time-correlation function of the pressure tensor, a viscosity indicator, through the Stokes-Einstein equation's analytic continuation into the Laplace domain. We integrate this equation with molecular dynamics (MD) simulations to derive necessary parameters. Our approach computes NMR lineshapes using a generalized diffusion coefficient, accounting for temperature and confinement geometry. This method directly links the memory function with thermal transport parameters, facilitating accurate NMR signal computation for non-Markovian fluids in confined geometries.
\end{abstract}

\maketitle

\section{\label{sec:Intro}Introduction}
NMR spectroscopy stands as a preferred method for probing molecular self-diffusion. Utilizing pulsed-field gradient (PFG) NMR experiments, it is possible to analyze molecular dynamics within porous media. This is achieved by examining fluid motion, even in structurally complex geometries~\cite{CallaghanBook11,price2009nmr,valiullin2016diffusion}. Such measurements provide insights into the microstructure of these media, elucidating characteristics like pore size, shape, and connectivity, as well as tortuosity.
However, recent experimental data from gaseous systems have indicated significant deviations from the predictions of conventional NMR theory. Contrary to the expected increase in line broadening at higher temperatures, empirical studies have reported a phenomenon of line narrowing~\cite{niknam2022nuclear,nature13,PRL15,PRLreb,JCP18}. These unexpected results have prompted a reevaluation of the existing theoretical framework to accurately account for these observations.

A new theory that yielded the correct temperature-dependent NMR trends~\cite{nature13,PRL15,PRLreb,JCP18}
was proposed based on a description of phase accumulation in diffusing nuclear spins whereby molecular dynamics are described by the generalized Langevin equation (GLE) with memory kernel.  The viscosity-dependent GLE-derived free induction decay expression captures the memory effects influenced by fluid dynamics in porous media, which are critical for understanding particle interactions within the medium and at  surfaces. The practical implications of measuring local viscosity are profound, enabling the investigation of intermolecular forces and surface characteristics in diverse systems such as geological substrates, catalytic processes, and biological matrices. Despite its utility, the GLE approach is challenged by boundary conditions, with restricted diffusion rendering the equation difficult to solve in closed form. To address this, MD simulations were proposed as a means to calculate effective viscosities in confined geometries~\cite{niknam2022nuclear}.

\begin{figure}[!htb]
\centering
\includegraphics[width=0.48\textwidth]{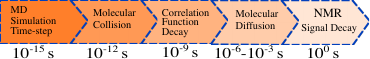}
\caption{This figure illustrates the timescales pertinent to fluid dynamics analysis via MD simulations. The interactions between fluid particles are captured with a time-step in the femtosecond (fs) range. Particle positions and velocities are sampled at 10-fs intervals, with correlation functions tracked until they decay, typically reaching times well into the hundreds of picoseconds (ps). For the decay of the NMR signal due to diffusion to be calculated, particle diffusion must be monitored over the course of hundreds of milliseconds (ms).}
\label{fig:timescale}
\end{figure}

MD simulations are useful for computing transport parameters such as viscosity in confined fluids, yet they face limitations with frequency-dependent phenomena~\cite{Starov01,Breugem07,Rudyak18,RizkPRL2022}. Specifically, high-frequency dynamics necessitate short computational time steps, thereby increasing computational demand. This is exemplified in our models of temperature-dependent NMR linewidths in gases, where we identified discrepancies between short-term velocity autocorrelation functions and long-term trajectory averages. The vast range of timescales (see Fig.~\ref{fig:timescale}), from femtosecond-level molecular collisions to the millisecond scale typical of NMR observations, presents a significant challenge for direct integration methods. These methods struggle to reconcile transient molecular memory effects with the prolonged timescales of NMR experiments, often resulting in data corrupted by computational noise. A better approach is needed to bridge this gap, and to ensure accurate representation of molecular dynamics within NMR linewidth predictions.

In prior work~\cite{niknam2022nuclear}, we developed a particle trajectory model that captures the temperature-dependent behaviors observed in free gas diffusion. This model integrates a Stochastic Differential Equation (SDE) for gas dynamics with MD simulations to accurately compute autocorrelation functions. This integrated approach allowed for the calculation of an effective viscosity via MD, which was then utilized within the SDE framework to simulate NMR lineshapes corresponding to experimental data. We termed this a two-scale methodology, with MD simulations detailing the microsecond ($\mu$s) dynamics and the SDE extending these insights to the timescale relevant to NMR measurements. This hybrid SDE-MD method effectively reconciles the time-scale gap between rapid molecular motions and the slower NMR observations.  

The problem of boundaries, however, remains unsolved.  In this study, we examine the connection between viscosity and memory kernel in the context of the GLE for NMR lineshape analysis subject to boundary conditions. Viscosity, a key transport property, encapsulates the memory effects crucial for investigating the properties of porous media.  The concept of ``effective viscosity'' has been used to describe transport in porous media~\cite{martys1994computer,song2021local,valdes2007effective,almalki2016investigations,kudenatti2018modelling,breugem2007effective,eberhard2019determination}.   Herein we demonstrate that viscosity can be deduced from the pressure tensor's correlation function, enabling the determination of memory functions to predict the NMR signal in the presence of boundaries. By applying the analytic continuation of the Stokes-Einstein equation to the Laplace domain, we derive a frequency-dependent diffusion function. This function is subsequently employed to calculate NMR lineshapes that are in agreement with experiments.

 \section{\label{sec:GLE} Generalized Langevin Equation:  Review}
 
Consider the Brownian motion of a spherical particle under the influence of forces that include a frictional component proportional to its velocity, denoted as $-\xi v$, as well as a stochastic force. The particle's motion is typically described by the Langevin equation, which assumes no memory of past velocities (Markovian dynamics):
\begin{equation*}
M \frac{ \mathrm{d}\bm{v}}{ \mathrm{d}t} = -\xi \bm{v} + \bm{\eta}_f(t),
\end{equation*}
where $M$ is the mass of the particle, $\bm{v}(t) = \dot{\bm{x}}(t)$ is its velocity, and $\bm{\eta}_f(t)$ represents a time-dependent stochastic force with finite variance and zero mean. In this model, the frictional force is proportional to the instantaneous velocity, and the stochastic force is uncorrelated at different times, characterized by a delta-function correlation, commonly referred to as white noise.

However, in many physical systems, the dynamics are non-Markovian, implying that the stochastic force exhibits  colored noise behavior and  the friction at time $t$ isn't solely dependent on the instantaneous velocity but is influenced by the trajectory of the particle's velocity. Consequently, friction is described by a memory function, frequently denoted as $\bm{\Gamma}(t)$, and integrated over the past time. This modification results in a GLE, which serves as an essential framework for describing the dynamics of viscoelastic fluids~\cite{PRL15}
\begin{equation}\label{eq:GLE}
 M\Dot{\bm{v}}+\int_0^t\Gamma(t-\tau)\bm{v}(\tau) \, \mathrm{d} \tau=\bm{\eta}_f(t).
\end{equation}
The memory kernel, $\Gamma(t)$,  is convolved with the particle's velocity history to model the influence of the viscous dynamics on the frictional forces. Within the GLE framework, all thermodynamic parameters and correlation functions undergo modifications through the incorporation of the memory function. For instance, if we consider $C(t)$ to represent a normalized correlation function pertaining to a parameter $A$, where $\langle A\rangle=0$, then $C(t)$ obeys the following integro-differential equation:
 \begin{equation}
 \dfrac{\mathrm{d}C}{\mathrm{d}t}=-\int_0^t \Gamma(\tau) C(t-\tau) \mathrm{d}\tau
 \end{equation}
 where the memory kernel is given by 
 \begin{equation}\label{eq:memform}
 \Gamma(t)=\langle \dot{A}(0) \exp[it(1-\mathcal{P})\mathcal{L}]\dot{A}(0)\rangle.
 \end{equation}
 In this equation, $\mathcal{L}$ is the Liouville operator describing the time evolution of $A(t)=\exp(i\mathcal{L}t )A(0)$, and $\mathcal{P}$ is a projection operator over the relevant variables~\cite{zwanzig2001nonequilibrium,mcq76}.  
 Consider the correlation function 
 \begin{equation}	
 C_{\eta}(\mathbf{k},t)=\frac{\langle J^*_{\perp}(\mathbf{k},0)J_{\perp}(\mathbf{k},t)\rangle}{\langle J^*_{\perp}(\mathbf{k},0)J_{\perp}(\mathbf{k},0)\rangle}
 \end{equation}
 where $\mathbf{k}$ is the wave number (normally taken along $\hat{z}$), and $\bm{J}(\mathbf{k},t)=\sum_{j=1}^N \dot{\bm{x}}_j(t)\text{e}^{ikz_j(t)}$ is the Fourier transform of the velocity, and $J_{\perp}$ is used to describe one of the two transverse components, assuming the second transverse component is zero. This is a normalized autocorrelation function that satisfies the equation
 \begin{equation*}
 \frac{\partial C_{\eta}}{\partial t}=-\int_0^t \Gamma_{\perp}(\mathbf{k},t)C_{\eta}(\mathbf{k},t-\tau) \,  \mathrm{d}\tau
 \end{equation*} 
 where $\Gamma_{\perp}(\mathbf{k},t)$ is given by Eq.~(\ref{eq:memform}) when 
 \begin{equation*}
 A=\frac{J_{\perp}(\mathbf{k},t)}{\langle J_{\perp}(\mathbf{k})J_{\perp}(\mathbf{k})\rangle^{1/2}}.
 \end{equation*}
 By solving this equation in the Laplace domain in the neighborhood of $s\rightarrow0$ and $\mathbf{k}\rightarrow0$ one finds the frequency dependent shear viscosity~\cite{berne1971time}
 \begin{equation}\label{eq:viscTCP}
 \eta(\omega)=\frac{1}{Vk_BT}\int_0^\infty \text{e}^{-i\omega t}\langle \bm{J}(0)\cdot\bm{J}(t)\rangle \, \mathrm{d}t
 \end{equation}
 where $V$ is the volume and $T$ is the temperature. At $\omega=0$  this corresponds to the equilibrium time correlation function, also known as the Green-Kubo autocorrelation function approach to deriving shear viscosity ~\cite{mcq76,rapaport_2004,todd_daivis_2017,Brush62,RizkPRL2022}
\begin{equation}
    \eta= \lim_{t\rightarrow\infty} \eta_{GK}(t)
\label{eq:GKintegral}
\end{equation}
with
\begin{equation}\label{eq:etaGK}
\eta_{GK}(t)= \frac{V}{3 k_B T}\int_0^{t} \sum_{\alpha < \beta} C_{\alpha \beta }(\tau) \, \mathrm{d}\tau,
\end{equation}
where $\alpha,\beta \in \{x,y,z\}$, $V$ is the volume and $T$ is the temperature. $C_{\alpha\beta}(\tau)= \left< p_{\alpha\beta}(\tau) p_{\alpha\beta}(0) \right>$ is the autocorrelation function of non-diagonal elements of the pressure tensor, e.g.
\begin{equation*}
p_{xy}(t)=\frac{1}{V}\Bigl\{\sum_j m_j v_{jx}(t) v_{jy}(t)+\frac{1}{2} \sum_{i\neq j} r_{ijx}(t) f_{ijy}(t)\Bigr\}.
\end{equation*}
Here $f_{ijy}$ represents the $y$-component of the force between two particles $i$ and $j$. The first term on the right hand side, is the kinetic contribution to the pressure tensor while the second term indicates the contribution from the potential energy. Other components $p_{\alpha\beta}$ of the pressure tensor are defined analogously.

The memory function formalism can be used to describe self-diffusion, denoting the probability of a particle being found in the volume $\mathrm{d}^3\mathbf{r}$ at time $T$ as $G(\mathbf{r},t)\mathrm{d}^3\mathbf{r}$, the diffusion equation is:
\begin{equation}\label{eqn:diffusion1}
\frac{\partial G(\mathbf{r},t)}{\partial t}=D\nabla^2 G(\mathbf{r},t)
\end{equation} 
where $D$ is the diffusion coefficient. The Fourier transform of the probability function $G(\mathbf{r},t)$ is given by $F_s(\mathbf{k},t)=\int_\mathbb{R} \text{e}^{i\mathbf{k} \cdot \mathbf{r}}G(\mathbf{r},t) \mathrm{d}^3\mathbf{r}$ 
and it also adheres to the memory function formalism through
\begin{equation}
\frac{\partial F_s(\mathbf{k},t)}{\partial t}=-\int_0^t \Gamma(\mathbf{k},t)F_s(\mathbf{k},t-\tau) \, \mathrm{d} \tau ,
\end{equation}
where $\Gamma(\mathbf{k},t)$ is calculated from Eq.~(\ref{eq:memform}) with $A=\exp(i\mathbf{k} \cdot \mathbf{r})$. 
 Fourier transformation of both sides of Eq.~(\ref{eqn:diffusion1})  gives 
\[\frac{\partial F_s(\mathbf{k},t)}{\partial t}=-k^2 DF_s(\mathbf{k},t) \]
and we find $F_s(\mathbf{k},t) \propto \exp(-\mathbf{k}^2 Dt) + f(\mathbf{k})$. The second derivative at the origin is $\bigl(  \frac{\partial ^2 F_s}{\partial k^2}\bigr)_{k=0}=-2Dt$, and we have 
\[\biggl( \frac{\partial ^2 F_s}{\partial k^2}\biggr)_{k=0}= -\frac{1}{3} \int_0^{\infty} 4\pi r^4 G( \bm{r},t) \, \mathrm{d}^3\mathbf{r} =-\frac{1}{3} \langle r^2(t)\rangle .\] 
So the average square displacement as a function time is found to be $\langle r^2(t)\rangle = 6Dt$,  which was originally derived by Einstein~\cite{einstein1905motion}. For a general starting point we can write this equation as
\begin{equation}\label{eqn:Ein} 
\langle |\bm{r}(t)-\bm{r}(0)|^2\rangle=6Dt
\end{equation}
which shows that the $G(\bm{r},t)$ can also be interpreted as the fraction of particles that end up in $\mathrm{d}^3\bm{r}$ given their various initial positions.

Again since $F_s$ in its general form is a complex-frequency dependent function, a frequency dependent diffusion function is defined as ~\cite{mcq76}
\begin{equation}\label{eq:Dw}
\mathcal{D}(\omega)=\frac{1}{3}\int_0^\infty \text{e}^{-i\omega t}\langle \bm{v}(0)\cdot \bm{v}(t)\rangle \, \mathrm{d}t. 
\end{equation}
The frequency-dependent diffusion function can also be defined using the generalized Fick's law, where the components of  current density are proportional to the gradient  of the concentration $\nabla c$ of the diffusing species, $j(\mathbf{k},\omega)=-\mathcal{D}(\mathbf{k},\omega)(\nabla c)_{\mathbf{k},\omega}$~\cite{zwanzig1964incoherent}.

The fluctuation-dissipation theorem, in its general form, can elucidate the nature of colored noise as represented by $\langle\bm{\eta}_f(0)\cdot\bm{\eta}_f(t)\rangle=k_BT\Gamma(t)$. This equation offers insight into the memory function, which corresponds to the correlation function of the stochastic force.  In its general form, a memory function is a wave-number and complex-frequency dependent function that describes the response of fluid, to fluctuations at different frequencies and at different length scales. Viewed through this lens, it is not unreasonable to anticipate that the correlation functions of various operators, such as the velocity autocorrelation function, viscosity, or diffusion coefficients, are subject to the same underlying decaying dynamics. In the subsequent section, we expound upon how this fundamental understanding can be harnessed to establish a connection between the GLE solution and the evaluation of the shear viscosity coefficient through MD simulations.

 \section{\label{sec:GSE} Generalized Stokes-Einstein equation}
 
The diffusion of a spherical particle in a purely viscous fluid at low Reynolds number is described by the Stokes-Einstein equation $D=\frac{k_BT}{6\pi a\eta}$, where $a$ represents the radius of the diffusing particle, and $\eta$ denotes the shear viscosity of the surrounding fluid at zero frequency. Consequently, the mean-square displacement of the diffusing particles increases linearly with time. We postulate a generalized form of the Stokes-Einstein equation by extending it to the frequency domain, introducing frequency dependence to its parameters. Beginning with the GLE equation (refer to Eq.~\ref{eq:GLE}), we apply a Laplace transformation, converting the convolution operator into a multiplication. The resultant equation is
 \begin{equation}
 \langle \bm{v}(0)\cdot \tilde{\bm{v}}(s)\rangle=\frac{k_BT}{Ms+\tilde{\Gamma}(s)}
 \end{equation}
where $s$ is the Laplace domain variable and parameters denoted with a tilde are assumed to be extended to the Laplace domain via analytic continuation. The equipartition theorem is often used to equate $M\langle \bm{v}^2(0) \rangle=k_BT$. The first term in the denominator represents the inertial effects and is negligible at low frequencies.   For the left-hand-side we can use the mean-square displacement instead of the velocity autocorrelation function. With $\bm{r}(t)-\bm{r}(0)=\int_0^t  \bm{v}(t') \, \mathrm{d}t'$ we have 
\[ [\bm{r}(t)-\bm{r}(0)]^2=\int_0^t  \int_0^{t}  \bm{v}(t')\cdot \bm{v}(t'') \, \mathrm{d}t' \, \mathrm{d}t''. \]
Averaging over an ensemble of particles we get:
\[ \langle [\bm{r}(t)-\bm{r}(0)]^2 \rangle =\int_0^t   \int_0^{t}   \langle \bm{v}(t')\cdot \bm{v}(t'') \rangle \, \mathrm{d}t' \, \mathrm{d}t''. \]
We set $\tau=t''-t'$ and change the limits of integration:
\begin{equation}\label{eqn:rsq}
 \langle [\bm{r}(t)-\bm{r}(0)]^2 \rangle = 2t \int_0^t  (1-\frac{\tau}{t})  \langle \bm{v}(0)\cdot \bm{v}(\tau)  \rangle  \, \mathrm{d}\tau
 \end{equation}
assuming that the equilibrium state is a stationary state with time reversal symmetry, as is the case in deterministic classical mechanics. For a fast decaying velocity autocorrelation function and long observation times this equation is approximated by
\begin{equation}\label{eqn:rsqapx}
 \langle \Delta r^2(t)\rangle=\langle [\bm{r}(t)-\bm{r}(0)]^2 \rangle \approx 2t\int_0^t \langle \bm{v}(0)\cdot \bm{v}(\tau) \rangle \, \mathrm{d}\tau.
 \end{equation}
Considering Eq.~(\ref{eqn:Ein}) the left hand side can be expressed in terms of the Laplace transform of the mean-square displacement,
 \begin{equation}\label{eq:Gammas}
 \tilde{\Gamma}(s)=\frac{6k_BT}{s^2\langle \Delta\tilde{r}^2(s)\rangle}.
 \end{equation}
The final step consists of relating the memory function to the viscosity. Mason et. al.~\cite{mason1997diffusing} achieve this by writing the following relation between macroscopic stress $\tau(t)$ and the stress relaxation modulus $G_\mathbf{r}(t)$
\begin{equation}
\tau(t)=\int_0^t  G_{\mathbf{r}}(t-t')\dot{\gamma}(t') \,  \mathrm{d}t'
\end{equation}
 where $\dot{\gamma}(t')$ is the strain rate. They noticed that the Laplace transform of $G_{\mathbf{r}}(t)$ had units of viscosity and by extending the zero-frequency version of Stokes law they arrived at
\begin{equation}\label{eq:Stokesw}
\tilde{G}_{\mathbf{r}}(s) =\frac{\tilde{\Gamma}(s)}{6\pi a}.
\end{equation} 
This relationship, which has been termed the ``correspondence principle'', provides a direct link between memory kernel and viscosity~\cite{cordoba2012elimination,zwanzig1970hydrodynamic,mason1995optical,mason2000estimating}.
 By combining Eqs.~(\ref{eq:Gammas}) and (\ref{eq:Stokesw}) we obtain a direct link between viscoelastic modulus $\tilde{G}(s)=s\tilde{G}_{\mathbf{r}}(s)$, and mean square displacement, which is a generalized form of the Stokes-Einstein equation
\begin{equation}
\tilde{G}(s)\approx \frac{k_BT}{\pi a s \langle \Delta \tilde{r}^2(s)\rangle}.
\end{equation} 
Considering the complex viscosity spectrum $\tilde{\eta}(s)= \tilde{G}(s)/s$  we arrive at
\begin{equation}\label{eq:GSE}
\tilde{\mathcal{D}}(s)=\frac{k_BT}{6\pi a s \tilde{\eta}(s)}.
\end{equation} 
 This equation was experimentally verified using diffusing-wave spectroscopy~\cite{mason1997diffusing}.

\section{\label{sec:PFGS}Pulsed Field Gradient (PFG) NMR}

The PFG NMR technique plays an important role in detecting molecular self-diffusion processes at the nanometer scale. This experimental method operates by assigning a spatial label to individual molecules based on their Larmor frequency, in the presence of a constant magnetic field gradient applied across the sample. The foundational concept, initially proposed by Hahn~\cite{HahnEcho}, exploits the precessional displacement of the signal phase, commonly referred to as signal dephasing (i.e., loss of phase coherence among the spins), within a spin echo experiment to quantify molecular translational motion. The theoretical foundation of such experiments was developed by Carr and Purcell~\cite{CP54}, and subsequently underwent further improvements~\cite{MG58,Stejskal65,callaghan1984pulsed}.

In its simplest form, the PFG NMR method is based on a Hahn echo sequence, denoted as $90-\tau-180-\tau$, where effects of spin-spin interactions are removed and  diffusion effects are studied under a constant magnetic field gradient $\bm{g}$, applied both before and after the echo pulse. The lineshape of NMR signal in the diffusion experiment is evaluated by a signal attenuation function $R(\tau)$ that describes the total signal from an ensemble of spins dephasing in a gradient. The cumulant expansion gives:
\begin{eqnarray}
R(\tau) &=& \left< \exp(i\int_0^\tau \omega(t)\mathrm{d}t) \right> \nonumber \\
&=& \exp\left(i\int_0^\tau\langle\omega(t)\rangle\mathrm{d}t -\frac{1}{2}\int_0^\tau\mathrm{d}t\int_0^\tau\mathrm{d}t' \langle\omega(t)\omega(t')\rangle + \cdots\right) \nonumber 
\end{eqnarray}
where frequency depends on position $\omega(t)=\gamma\bm{g}\cdot\bm{x}(t)$. For a stationary process, the ensemble average in the first term is constant. For the second term
\begin{equation}\nonumber
\int_0^\tau \mathrm{d}t \int_0^\tau \mathrm{d}t' \langle x(t) x(t') \rangle = 2\int_0^\tau (\tau-t) \langle x(t)x(0) \rangle \mathrm{d}t.
\end{equation}
We extend the upper limit of the integral to infinity and invoke the Plancherel's theorem for Laplace transforms of real-valued functions $f(t)$ and $g(t)$:
\[
\int_0^\infty f(t) g(t) \, dt = \frac{1}{2\pi i} \int_{c - i\infty}^{c + i\infty} \tilde{f}(s) \tilde{g}(-s) \, \mathrm{d}s.
\]
where the Bromwich contour has the usual meaning (vertical line with real part fixed, and placed to the right of all poles).
We get:
\begin{eqnarray} &&\int_0^\infty (\tau-t) \langle x(t)x(0) \rangle \mathrm{d}t =\nonumber \\ 
&&\frac{1}{2\pi i}  \int_{c - i\infty}^{c + i\infty}   \widetilde{(\tau-t)}(s) \widetilde{ \langle x(t)x(0) \rangle} (-s) \mathrm{d}s \nonumber \end{eqnarray}
Let us abbreviate:
$$ X(s) 
 \equiv \widetilde{ \langle x(t)x(0) \rangle} (s) $$
$$ U(s) \equiv  \widetilde{  (\tau-t) } (s)   = \frac{\tau}{s} - \frac{1}{s^2} $$
so that
\begin{eqnarray}
 &&\int_0^\infty (\tau-t) \langle x(t)x(0) \rangle \mathrm{d}t = \int_{c - i\infty}^{c + i\infty}  U(s) X(-s) \mathrm{d}s  \nonumber\\
 &=& \int_{c - i\infty}^{c + i\infty}  U(s) \frac{  V(-s) }{ s^2 }  \mathrm{d}s    = \int_{c - i\infty}^{c + i\infty}  \frac{U(s) }{ s^2 }  V(-s)   \mathrm{d}s  \nonumber
 \end{eqnarray}
where the second equality follows from $V(s) = s^2 X(s)$.  Another application of the Plancherel theorem (inverse Laplace transform) yields:
$$ \int_0^\infty  \left(  \tau  \frac{ t^2}{ 2! } - \frac{t^3}{ 3!} \right) \langle v(t) v(0) \rangle  \mathrm{d}t  \approx M \tau \mathcal{T}^2 \int_0^\tau   \langle v(t) v(0) \rangle  \mathrm{d}t $$
where $\tau \gg \mathcal{T}$ is assumed, employing the intermediate value theorem with the constant $M=O(1)$, and considering a time factor $\tau_c \le \mathcal{T} \le \tau$ bounded from below by the correlation time $\tau_c$ and the form of the velocity autocorrelation function $\langle v(t) v(0) \rangle$. For small molecules in weak gradient fields, the cumulant expansion is typically truncated at the second order~\cite{stepivsnik1999validity}. In isotropic diffusion, after the decay of the velocity correlation function, the diffusion function $\mathcal{D}(t)=\frac{1}{3}\int_0^t \langle \mathbf{v}(t')\cdot\mathbf{v}(0)\rangle\mathrm{d}t'$ becomes constant. The signal attenuation due to the second cumulant term, is:
\begin{equation}\label{eq:decayfac}
R(\tau) = \exp(-3 M \gamma^2g^2 D \tau \mathcal{T}^2).
\end{equation}
Accurately calculating frequency-dependent viscosity or diffusion coefficients remains challenging. The subsequent section illustrates how MD simulations can effectively model fluid dynamics to assist in this endeavor.

\section{\label{sec:MDiso} Molecular Dynamic Simulation of Signal Attenuation}

MD simulations are suitable for generating the autocorrelation functions necessary to evaluate frequency-dependent diffusion and viscosity parameters. Our model must accurately capture the complex diffusion dynamics exhibited by gas particles, which display memory effects~\cite{nature13,PRL15,PRLreb,JCP18,niknam2022nuclear}. This is the appropriate description for non-Markovian liquids, non-ideal gases, and intricate geometries where assessing transport properties poses considerable challenges.

The process involves computation of interatomic or intermolecular forces and the subsequent motion of neighboring particles. The list of neighboring particles, which is crucial for calculating interactions, is dynamically updated at each time step.  The simulation boundaries can be periodic, allowing particles leaving the simulation box to re-enter it from the opposite side.  (Periodic boundary conditions are a common technique to mimic an infinite system.) At predefined time intervals, physical attributes of particles, such as position and momentum, are sampled to compute averages, which are recorded in an output data file~\cite{FRENKEL200263,rapaport_2004}. To facilitate the generation of these parameters for a particle ensemble, we employed the software ``Large-scale Atomic/Molecular Massively Parallel Simulator'' (LAMMPS), a popular open-source computational tool~\cite{LAMMPS} for MD simulations.

MD simulations were conducted to calculate the shear viscosity of gaseous xenon (Xe), as defined by Eq.~(\ref{eq:etaGK}). We utilized the Lennard-Jones (LJ) pair potential, expressed as $U(r) = 4\epsilon[(\sigma/r)^{12} -(\sigma/r)^6]$, where interactions between xenon atoms were characterized by $\epsilon = 1.77$ kJ/mol, the depth of the potential well, and $\sigma=4.1$ \AA, the distance at which the potential energy becomes zero. Our simulations of bulk fluid were conducted for isotropic diffusion by placing 2,000 xenon atoms within a box defined by periodic boundary conditions. Throughout the simulations, we maintained a consistent particle count, volume, and temperature, adhering to the canonical ensemble ($NVT$ ensemble)~\cite{niknam2022nuclear}. Each set of simulations was repeated for 10 different random seeds for the initial positions and velocities of the particles to ensure robust statistical sampling and accuracy of the results. In the simulations of restricted diffusion (i.e., diffusion limited by the nanotube geometry), nanotubes of a fixed length and various diameters were employed, and the number of particles was adjusted to maintain a constant particle density.

\section{\label{sec:GasB} Simulation of Bulk Gas}

In the first example, we investigated the unrestricted self-diffusion of particles in a gaseous phase, a phenomenon that has been observed to exhibit an unexpected line-narrowing effect with increasing temperature~\cite{nature13,PRL15,PRLreb,JCP18,niknam2022nuclear}. Simulations were carried out across a temperature range spanning from 200~K to 400~K, encompassing a total of 21 distinct temperature values.  To ensure equilibration, particle trajectories were evolved for a duration of $10^6$ fs. During each simulation, the $C_{\alpha \beta}$ correlation functions were computed 60 times, extending data collection to ensure the decay of correlations is fully captured. Subsequently, these correlation functions were averaged and utilized to compute the Green-Kubo integral, Eq.~\eqref{eq:GKintegral}. The effectiveness of this approach relies on the correlation functions exhibiting similar decay rates, ensuring the integrals converge to a representative average value.

\begin{figure}[!htb]
\centering
\includegraphics[width=0.40\textwidth]{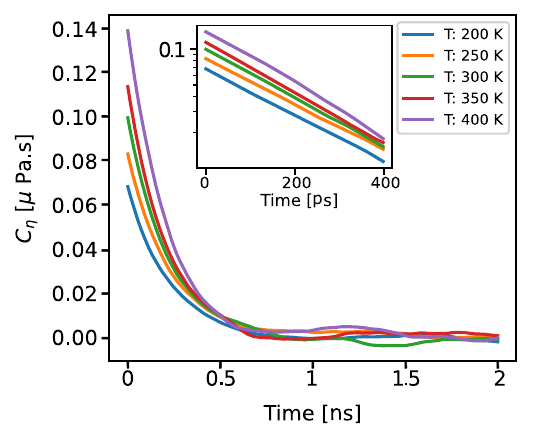}
\caption{The autocorrelation function of the pressure tensor, as specified in Eq.~(\ref{eq:etaGK}), was calculated via MD simulation and plotted as a function of time for a series of temperatures. The inset provides a semi-logarithmic view of the initial data points, illustrating that the autocorrelation function follows a clear exponential decay with a consistent rate across different temperatures. The viscosity coefficient, obtained from the integral of the autocorrelation function values, correlates directly with the amplitude of the initial point on the decay curve.}
\label{fig:Cmugas}
\end{figure}

Figure~\ref{fig:Cmugas} illustrates the temporal decay behavior of the correlation of the pressure tensor, $C_{\eta}=\frac{1}{3}(C_{xy}+C_{yz}+C_{yz})$, at various temperatures. The inset of Figure~\ref{fig:Cmugas} provides a semi-logarithmic plot of the initial segment of this data.
We observe that the rate of decay, indicated by the consistent slope in the semi-logarithmic plot, remains consistent across all temperatures. Therefore, variations in the diffusion coefficient can be attributed to differences in the initial amplitude of the pressure tensor correlation function, as shown in Eq.~(\ref{eq:etaGK}). The decay of the pressure correlation function can be described with:
\begin{equation}\label{eq:Cmu}
C_{\mu} \approx A \exp(-t/T_D)
\end{equation}
where  $A$ represents the temperature-dependent amplitude and $T_D$ is the common decay characteristic time due to diffusion.

\begin{figure}[!htb]
\centering
\includegraphics[width=0.40\textwidth]{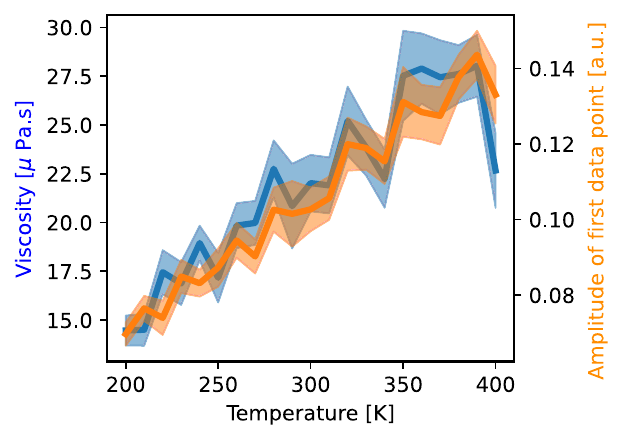}
\caption{The blue data points represent the viscosity coefficient obtained by numerically integrating the pressure correlation function at various temperatures. The consistent decay rate in the correlation function suggests that the amplitude of the initial data point provides an accurate and reliable approximation of the viscosity coefficient, as indicated by the orange data points.}
\label{fig:CmuInt_gas}
\end{figure}

It is essential for the initial amplitude of the correlation function decay curve to align with its integral, indicative of the viscosity factor. Figure~\ref{fig:CmuInt_gas} illustrates these parameters across various simulation temperatures. Our previous work demonstrated that our simulated viscosity parameters conform to Sutherland's formula, which describes temperature-dependent variations in viscosity~\cite{niknam2022nuclear}. 
\begin{equation}
\eta \propto \frac{T^{3/2}}{T + C}
\end{equation}
Here, $C$ represents Sutherland's constant. Correspondingly, the initial value of the $C_{\alpha\beta}$ correlation function follows a similar pattern. To derive the generalized diffusion function $\tilde{\mathcal{D}}(s)$ from the complex-frequency-dependent viscosity function within the Laplace domain, $\tilde{\eta}(s)$, we approximate $C_{\mu}$ using an exponential decay function:
\begin{equation}
\tilde{\eta}(s) = \widetilde{A\text{e}^{-t/T_D}} (s) = \frac{A}{T_D^{-1} + s}.
\end{equation}
In the Laplace domain, we link the viscosity function to the diffusion function using the generalized Stokes-Einstein equation, Eq.~(\ref{eq:GSE}), 
\begin{equation}
\tilde{\mathcal{D}}(s) = \frac{k_BT}{6\pi a A s (T_D^{-1} + s)}.
\end{equation}
With an inverse Laplace transform we find a generalized diffusion coefficient in the time domain
\begin{equation}\label{eq:Dt}
\mathcal{D}(t)=\frac{k_BT}{6\pi a}\frac{\theta(t)}{AT_D}
\end{equation} 
where $\theta(t)$ is the Heaviside step function. So we recover the zero frequency Stokes-Einstein relation where $\eta$ is replaced by $AT_D$. Thus, for diffusion in bulk gas, it is not necessary to consider the frequency-dependent diffusion coefficient, and the signal decay attributed to particle diffusion, as delineated in Eq.~(\ref{eq:decayfac}), can be conveniently addressed by invoking the zero-frequency Stokes-Einstein equation. This results in a line-width that is inversely proportional to the zero-frequency viscosity coefficient, as previously discussed in~\cite{niknam2022nuclear}
\begin{equation}\label{eq:lw0freq}
\Delta \omega=(3M\gamma^2g^2D)^{1/3}=(3M\gamma^2g^2\frac{k_B T}{6\pi a A T_D})^{1/3}.
\end{equation}

\begin{figure}[!htb]
\centering
\includegraphics[width=0.40\textwidth]{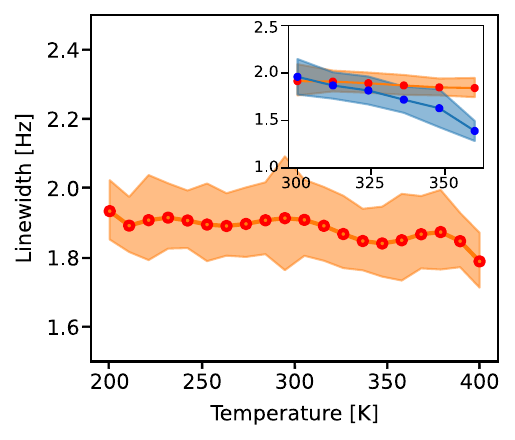}
\caption{Linewidth of the NMR signal calculated using Eq.~(\ref{eq:lw0freq}) and derived from the simulated viscosity coefficient. A line-narrowing effect is observed. The inset shows a comparison with experimental results.}
\label{fig:lwGasB}
\end{figure}

Utilizing this equation, the viscosity coefficients derived from MD simulations enable the prediction of gas-phase lineshape trends. Figure~\ref{fig:lwGasB} demonstrates this for a magnetic field gradient of $0.01$ G/cm, with the value of $M$ fitted to experimental results. The figure's inset offers a direct comparison between the simulated linewidth and the experimentally measured linewidth of gaseous tetramethylsilane (TMS). The experimental sample comprised a sealed tube containing liquid TMS, prepared using the freeze-pump-thaw method for degassing. The sample was then heated, elevating its temperature above $25^{\circ}$C, to promote TMS evaporation. Measurements were conducted using a Bruker AV 600 MHz NMR spectrometer, equipped with variable temperature control and pulsed gradient capabilities, ensuring a controlled, precise environment for data acquisition.

\section{\label{sec:MDaniso} Restricted Diffusion}

The analysis of anisotropic diffusion in NMR requires extending the zero-frequency diffusion coefficient to a diffusion tensor. In cases like aligned fibers, NMR signal attenuation significantly depends on the measurement direction. This dependence is determined by the orientation of the sample relative to the field gradient vector, represented by its Cartesian components $\bm{g}=(g_x, g_y, g_z)$. The NMR lineshape is a weighted average of all spins in an ensemble, while the signal decay rates are determined by the diffusion tensor's components.

For more intricate samples, such as those with multiple orientations or powder-like distributions, averaging over all possible orientations becomes essential. This method is akin to assuming a single oriented domain where the gradient direction is randomly sampled~\cite{callaghan1984pulsed,price2009nmr}.

In this study, we employed the memory function formalism to explore the complexities of restricted diffusion in fluids. Our series of MD simulations focused on the restricted diffusion of gas particles, specifically investigating the effects of temperature variations and different diameters of cylindrical boundaries on their diffusion behavior. These simulations were conducted inside cylinders of various diameters, all aligned parallel to the magnetic field gradient $\bm{g}=(0, 0, g)$. The objective was to understand how NMR linewidth varies with temperature and cylinder radius.

\begin{figure}[!htb]
\centering
\includegraphics[width=0.40\textwidth]{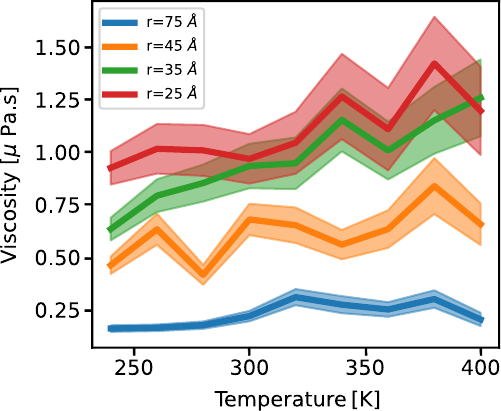}
\caption{Viscosity coefficient evaluated for cylinders of uniform length but varying diameters. This coefficient was ascertained by integrating all three components of the pressure tensor cross-correlation function. Notably, the temperature trend observed closely parallels that of bulk gas diffusion. Furthermore, the viscosity coefficient demonstrates a clear sensitivity to the diameter of the cylindrical tube.}
\label{fig:viscTubes}
\end{figure}

For the simulations, we used Xenon (Xe) particles. The interactions between the Xe particles and the cylindrical boundary were modeled using the Lennard-Jones potential, with parameters $\epsilon = 0.3$ kJ/mol and $\sigma = 4.295$ \AA, representing a silica tube. To determine the viscosity coefficient, we integrated all three components of the pressure tensor, referred to as $C_{\alpha\beta}(\tau)$ in Eq.~(\ref{eq:etaGK}). Although the $C_{xy}$ component showed significantly higher values than the $C_{yz}$ and $C_{xz}$ components, given the tube's orientation along the \(z\)-axis, we opted to integrate all three components together for a thorough analysis, as illustrated in Fig.~\ref{fig:viscTubes}.

The simulations indicate that viscosity in this context exhibits a temperature dependence akin to that observed in bulk gas. Notably, as the temperature rises, there is a marked increase in the medium's resistance to molecular motion, likely resulting from more frequent molecular collisions at elevated temperatures. Consequently, this increased resistance is expected to raise the diffusion coefficient, subsequently narrowing the NMR lineshape.

It is also significant to note that the viscosity observed here is substantially lower—by approximately an order of magnitude—than that in bulk fluid environments. This observation points to a markedly reduced resistance to molecular motion within the medium. Moreover, the simulations suggest an increase in resistance to diffusion when molecules are confined within tubes with smaller diameters.

\begin{figure}[h]
\centering
\includegraphics[width=0.48\textwidth]{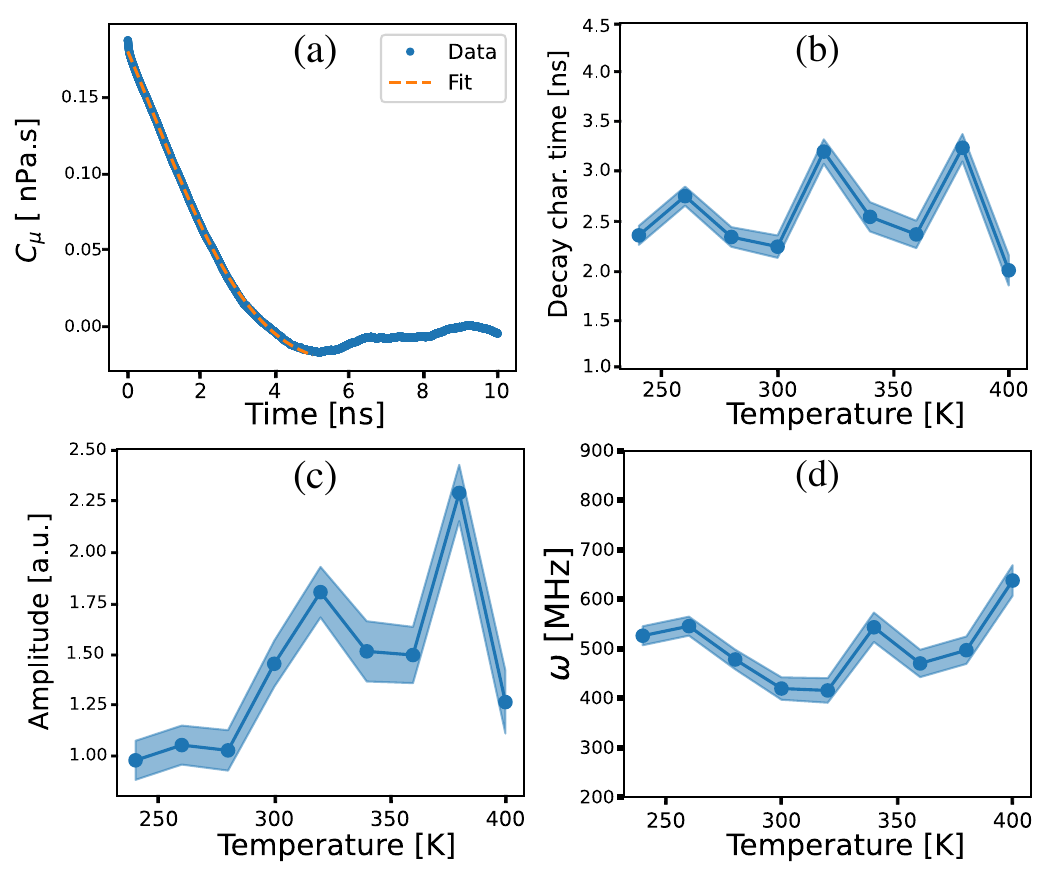}
\caption{Restricted diffusion of Xe gas molecules in a cylindrical tube with a radius of 75 \AA. Panel (a) displays the decay pattern of the gas molecules, highlighting the presence of an oscillatory component alongside exponential decay. Panel (b) depicts the variation in the characteristic decay time $T_D$ across different temperatures, illustrating a lack of a consistent trend. Panel (c) shows that the amplitude of the decay function escalates with an increase in temperature, aligning with observations in bulk gas behavior. Panel (d) presents the variation in the angular frequency of the oscillatory component with temperature.}
\label{fig:A75fit}
\end{figure}

To better comprehend the interplay between viscosity and the dynamics of complex fluid flow, we explored its correlation with the radius of the cylindrical boundary. In cases of restricted fluid diffusion within these cylindrical confines, the decay pattern of the pressure correlation function diverges from a simple exponential decay. This divergence is depicted in Fig.~\ref{fig:A75fit}a. A more precise representation of this decay pattern is obtained by integrating an exponential decay with an oscillatory component, modeled as:
\begin{equation}\label{eq:Cmu}
C_{\mu} \approx A \exp(-t/T_D) \cos(\omega t),
\end{equation}
where $\omega$ denotes the frequency of the oscillatory component.

Figure~\ref{fig:A75fit} presents the fitting results of the simulated pressure tensor data for a tube with a radius of 75~\AA, using Eq.~(\ref{eq:Cmu}). Our initial analysis reveals that the characteristic decay time, similar to that in bulk fluids, varies minimally with temperature. However, the amplitude of the decay curve exhibits significant temperature dependence, as shown in Figs.~\ref{fig:A75fit}b and c.

\begin{figure}[!htb]
\centering
\includegraphics[width=0.40\textwidth]{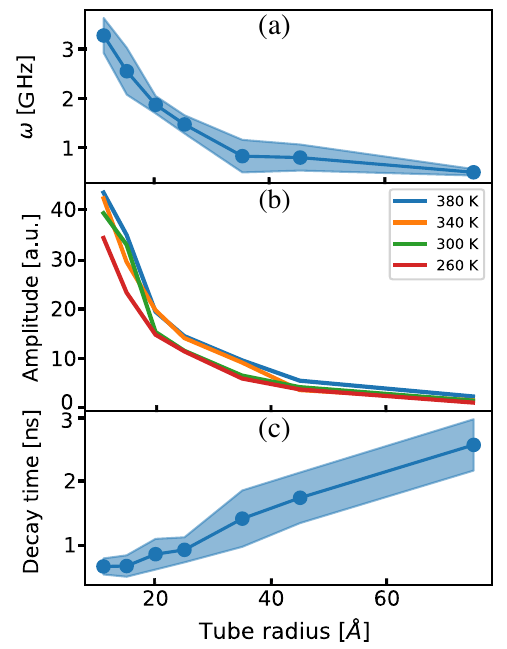}
\caption{The frequency of the oscillatory term and the amplitude of the pressure tensor decay exhibit well-defined relationships with the tube radius. Panel (a) shows that the average frequency decreases exponentially with the tube radius. Panel (b) illustrates a similar exponential relationship for amplitudes at a constant temperature. Panel (c) displays the variation in the characteristic decay time of the pressure tensor as a function of the tube radius.}
\label{fig:avgwM}
\end{figure}

The angular frequency of the oscillatory component remains relatively stable across different temperatures but shows a notable correlation with the tube's radius. Figure~\ref{fig:avgwM}a displays this oscillation frequency, averaged over various temperatures, plotted against the cylindrical tubes' radii. This pattern suggests that the oscillatory term's frequency acts as an indicator of diffusion dimensions, being proportional to the radius with an exponentially decaying function. This underscores the importance of viscosity in examining geometrical constraints affecting diffusion within the tubes.

The amplitude of the decay curve is influenced by both the temperature and the radius of the cylindrical tube. At a fixed temperature, we observe that the amplitudes exhibit an exponential correlation with the tube radii, as demonstrated in Fig.~\ref{fig:avgwM}b. The characteristic decay time of the pressure correlation function, similar to that observed in bulk diffusion, shows minimal variation with temperature. However, it noticeably increases as the tube radius expands, as clearly depicted in Fig.~\ref{fig:avgwM}c. It is important to note that, while both the frequency and amplitude of the pressure tensor decay curve show exponential dependence on the tube radius, their distinct impacts are more pronounced in experiments conducted across various temperatures.

\section{\label{sec:linewidth} Effects on NMR Linewidth}

The analysis from the preceding section shows that key parameters, such as the characteristic decay time, amplitude, and frequency associated with the oscillatory component of viscosity, exhibit consistent and predictable trends. The main objective is to correlate these parameters with the generalized diffusion coefficient and determine their impact on the NMR signal's lineshape. This aligns with the broader goal of using the NMR signal profile to deduce detailed characteristics of complex fluid behavior.

Our analysis begins with the generalized Stokes-Einstein equation (refer to Eq.~\ref{eq:GSE}). In this context, we deduce the generalized diffusion function $\tilde{\mathcal{D}}(s)$ from the complex-frequency-dependent viscosity function within the Laplace domain. We obtain the viscosity function by applying a Laplace transform to the time-domain viscosity expression, represented as $\tilde{\eta}(s)$. Utilizing analytical expressions to approximate the time-dependent pressure tensor $C_{\mu}$, we derive:
\begin{equation}
\tilde{\eta}(s)=\widetilde{ A\text{e}^{-t/T_D}\cos(\omega t) } (s) =\frac{A(1/T_D+s)}{\omega^2+(1/T_D+s)^2}.
\end{equation}
Employing the generalized Stokes-Einstein equation we get:
\begin{equation}
\tilde{\mathcal{D}}(s)=\frac{k_BT}{6\pi a}\frac{\omega^2(1/T_D+s)^2}{As(1/T_D+s)}.
\end{equation} 
Next we perform an inverse Laplace transform to get the generalized diffusion function in the time domain
\begin{equation}\label{eq:Dt}
\mathcal{D}(t)=\frac{k_BT}{6\pi a}\frac{\theta(t)}{A}[1/T_D+\omega^2 T_D(1-\text{e}^{-t/T_D})]
\end{equation} 
where $\theta(t)$ is the Heaviside step function.

\begin{figure}[!htb]
\centering
\includegraphics[width=0.40\textwidth]{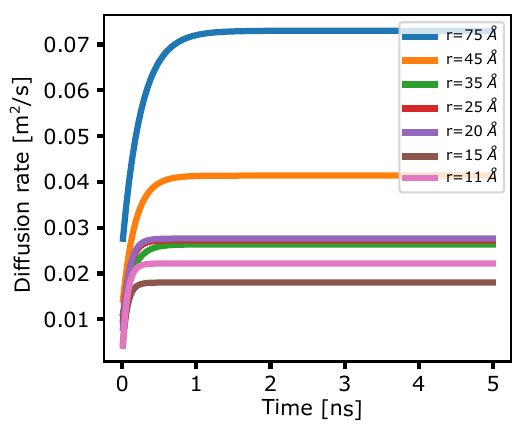}
\caption{Generalized diffusion parameter calculated according to Eq.~(\ref{eq:Dt})  at T=300 K. Transport parameters are extracted from MD simulation of Xe gas diffusion in cylindrical tubes. }
\label{fig:DtvsR}
\end{figure}

To more closely examine the impact of tube radius, we analyzed the diffusion function at a constant temperature. Figure~\ref{fig:DtvsR} illustrates the results for cylindrical nanotubes with radii ranging from 11 to 75~\AA. Initially, the diffusion function demonstrates time-dependent behavior, especially noticeable when the pressure tensor correlation function $C_{\mu}$ has not yet stabilized to zero. Beyond this phase, extending over several multiples of $T_D$ (the characteristic diffusion time), a consistent diffusion coefficient becomes apparent. The time dependency observed in our simulations reflects the time required for particles to reach and interact with the tube boundaries, representing a temporal dynamic arising from our observational methodology rather than an intrinsic property of the particle diffusion process itself. In reality, the diffusion function represents the mean square displacement averaged over all particles, which assumes a time-independent value. Thus, for diffusion analysis within the GLE formalism, it is appropriate to use the equilibrium value of $\mathcal{D}(t)$. As outlined in Eq.~(\ref{eq:Dt}), this equilibrium value is influenced by factors like $T_D$ and $\omega$, leading to diffusion coefficients that significantly differ from those in bulk systems and fluids without memory effects.

\begin{figure}[!htb]
\centering
\includegraphics[width=0.40\textwidth]{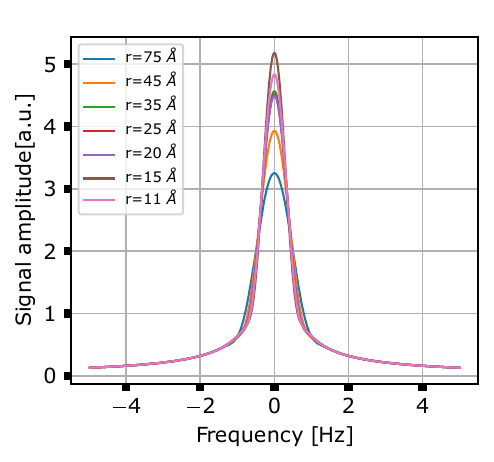}
\caption{The linewidth of NMR signal  at T=300 K is estimated for nanotubes with various pore sizes, using the generalized diffusion parameter. Larger diffusion parameters correspond to a broader line. }
\label{fig:NMRsignal300}
\end{figure}

The next step is to analyze the effect of the diffusion function on the NMR signal's lineshape. Once again, we utilize the attenuation function described in Eq.~(\ref{eq:decayfac}), fitting the value of $M$ and adopting the equilibrium value of the generalized diffusion function $\mathcal{D}$. Figure~\ref{fig:NMRsignal300} illustrates the NMR lineshape trend. The data reveals a distinct pattern: larger pore sizes correspond to increased diffusion parameters, resulting in broader NMR lines.

\section{Conclusion}

This work highlights the significant role of memory functions in understanding the non-Markovian dynamics of viscoelastic fluids. By employing frequency-dependent viscosity and diffusion coefficients, and building upon the generalized Stokes-Einstein equation, we have developed a generalized diffusion function. This function is notably sensitive to memory effects, as demonstrated by its relationship to the pressure correlation function. Most importantly, we introduce a novel method for analyzing NMR lineshape in complex geometries, using transport parameters derived from MD simulations.

\section{Acknowledgments}
The computational and storage services used in this work are part of the Hoffman2 Shared Cluster, provided by the UCLA Institute for Digital Research and Education’s Research Technology Group. L.-S. B. is grateful for the support from NSF award CHE-2002313 and for the valuable discussions with Thomas G. Mason regarding the scope of applicability of the generalized Stokes-Einstein equation.

\section{Data Availability Statement}

The data that support the findings of this study are available from the corresponding author upon reasonable request. 
Jupyter notebooks used in this study can be downloaded at:
\texttt{https://doi.org/10.5061/dryad.m905qfv7n}

\bibliographystyle{naturemag}
\bibliography{VACFMD.bib}

\end{document}